# Absolute efficiency calibration of $^6$LiF-based solid state thermal neutron detectors

Paolo Finocchiaro[1,*], Luigi Cosentino[1], Sergio Lo Meo[2], Ralf Nolte[3], Desiree Radeck[3]

1.  INFN, Laboratori Nazionali del Sud, Catania, Italy
2.  ENEA Bologna, Italy
3.  PTB Braunschweig, Germany

**Abstract**

The demand for new thermal neutron detectors as an alternative to $^3$He tubes in research, industrial, safety and homeland security applications, is growing. These needs have triggered research and development activities about new generations of thermal neutron detectors, characterized by reasonable efficiency and gamma rejection comparable to $^3$He tubes. In this paper we show the state of the art of a promising low-cost technique, based on commercial solid state silicon detectors coupled with thin neutron converter layers of $^6$LiF deposited onto carbon fiber substrates. A few configurations were studied with the GEANT4 simulation code, and then calibrated at the PTB Thermal Neutron Calibration Facility. The results show that the measured detection efficiency is well reproduced by the simulations, therefore validating the simulation tool in view of new designs. These neutron detectors have also been tested at neutron beam facilities like ISIS (Rutherford Appleton Laboratory, UK) and n_TOF (CERN) where a few samples are already in operation for beam flux and 2D profile measurements. Forthcoming applications are foreseen for the online monitoring of spent nuclear fuel casks in interim storage sites.

## 1 Introduction

The lack and the increasing cost of $^3$He have triggered in the last years a worldwide R&D program investigating new techniques for neutron detection. For many applications a realistic alternative is needed to $^3$He-based neutron detectors which so far have been the most widely used systems, as they are almost insensitive to radiation other than thermal neutrons [1],[2],[3].

Several developments involving neutron detection are currently being pursued in the fields of homeland security, nuclear safeguards, nuclear decommissioning and radwaste management. Two possible applications are worth mentioning, namely the development of neutron sensitive panels to be placed around nuclear material in a $\approx 4\pi$ solid angle coverage for coincidence neutron counting applications [4], and the deployment of arrays of small neutron detectors for the online monitoring of spent nuclear fuel storage sites [5],[6].

In a previous paper [7] it was shown that the use of a fully depleted silicon detector, coupled with a $^6$LiF neutron converter film deposited onto an independent substrate, can be successfully exploited to detect thermal neutrons with a reasonable efficiency. The neutron conversion mechanism is based on the well known reaction

$$^6Li + n \rightarrow {}^3H \ (2.73 \ MeV) + \alpha \ (2.05 \ MeV) \qquad (1)$$

which is the only possible decay channel following the neutron capture in $^6$Li, and is free of gamma rays. Its cross section at thermal neutron energy is 940 b, and it scales with 1/v up to $\approx 200$ keV with a back-to-back isotropic emission of the reaction products. In refs. [8] and [9] detailed justifications were given for the convenience of using $^6$LiF as neutron converter material on a planar detector, which can be summarized as follows:
- higher kinetic energy of the decay products following the neutron capture in $^6$Li, in comparison with $^{10}$B even though the cross section is four times lower;
- ease of fabrication and assembling;
- robustness;
- possible reusability of the silicon diode and of the converter as independent components;
- low cost;
- possible position sensitivity by using a single or double side silicon strip detector.

---

* corresponding author, e-mail FINOCCHIARO@LNS.INFN.IT

Moreover, the detection efficiency one can achieve with this configuration, i.e. separate converter and silicon diode, is lower than, but comparable with, what is obtained by depositing the converter directly onto a micro- or nano-structured silicon diode [10], while showing basically a 100% production yield even with 25 cm$^2$ area detectors.

The reliability of this technique, along with a characterization in terms of response, efficiency and gamma sensitivity, was also assessed by means of GEANT4 simulations [11]. The technique is indeed well established [12],[13], and several applications are already in use like for instance at the n-TOF spallation neutron beam facility [14],[15]. The energy spectrum measured by the silicon detector in such a configuration has a characteristic shape, and allows to discriminate the capture reaction products from the low-energy background basically due to gamma rays. Even though we found an excellent agreement between simulation and experimental data taken with neutron sources and beams, it only concerned the spectrum shape as in all the tests performed there was only a rough knowledge of the incoming thermal neutron flux. In most applications of neutron detectors a quantitative assessment of the measured neutron flux is required, hence the need to determine the absolute efficiency of the detectors and to verify the reliability of the simulations.

In this paper we report on the efficiency calibration of a few samples of this solid state neutron detector making use of a certified thermal neutron field, and the results are compared to the respective GEANT4 simulations.

## 2  Experimental setup

Neutrons cannot be constrained by means of electromagnetic fields, and they only interact with matter via elastic and nuclear reactions. Predicting their trajectories, doses and fluxes is not straightforward and needs numerical simulations, in particular for thermal neutrons. While the efficiency of charged particle and gamma detectors can be easily measured by means of standard laboratory sources, this is not the case with neutron detectors. In order to measure the intrinsic efficiency of a thermal neutron detector one can: (i) place it at a given position in a generic neutron field and compare the number of counts per unit area with the same quantity as measured by a reference detector placed in the same position (relative efficiency calibration); (ii) place it in a well-known reference neutron field and calculate the ratio of the detected counts to the real number of impinging neutrons (absolute efficiency calibration).

### 2.1  The thermal neutron field

For the efficiency calibration of the devices we used a thermal neutron field at the Physikalisch-Technische Bundesanstalt (PTB). The Thermal Neutron Calibration Facility at PTB ([16],[17]) consists of sixteen $^{241}$Am-Be sources that are mounted inside a graphite block whose dimensions are 150 cm (height), 150 cm (width), and 180 cm (depth). The reference position is at 30 cm from the front surface of the moderator exit window and 75 cm above the floor. The neutron and photon fields at the reference position were characterized by means of measurements and Monte Carlo simulations [17]. The neutron field is highly thermalized with Maxwellian distribution, 98.4 % of neutrons have energies below the cadmium cut-off energy with a thermal neutron flux at the reference position of 68.3±1.9 neutrons/cm$^2$/s and a uniform field size of at least 10 x 10 cm$^2$.

Optionally, a cadmium plate can be installed in front of the moderator exit window that cuts the thermal neutron contribution below the cadmium threshold. This way, a pure thermal field can be obtained by applying the difference method. However, the high-energy (Eγ ≥ 5MeV) gamma ray flux increases by about one order of magnitude when the cadmium plate is inserted.

*2.2 The thermal neutron detectors*

The thermal neutron detectors we discuss in this paper were named SiLiF, because they feature silicon detectors (3 cm x 3 cm x 300 μm) and $^6$LiF converters. Two samples were calibrated at PTB (Fig. 1):

- A SiLiF1.6 made of a 1.6 μm $^6$LiF layer (thin converter) deposited onto a carbon fiber plate (a), coupled to a 300 μm thick silicon detector (b), and enclosed in a thin aluminum box (d).
- A SiLiF64 made of a double sandwich of four 16 μm $^6$LiF layers (thick converters) deposited onto carbon fiber plates (c), and enclosed in a thin aluminum box (d).

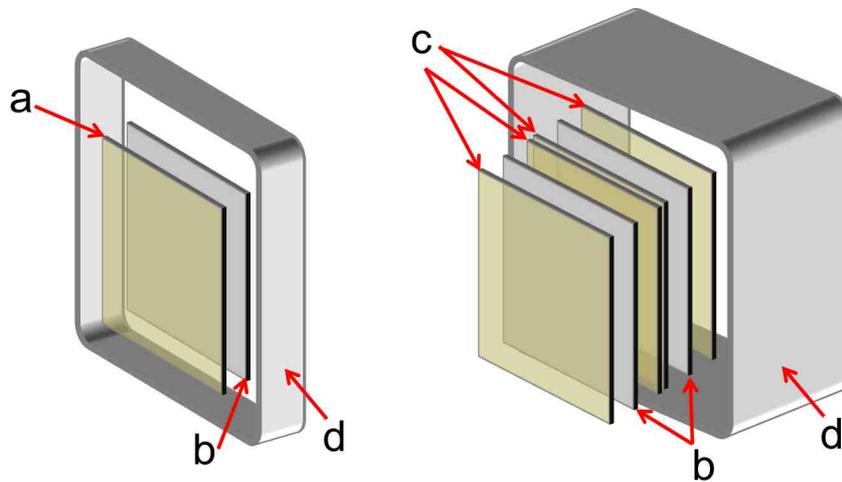

Fig. 1. The two thermal neutron detector samples calibrated at PTB. Left: SiLiF1.6 made of a 1.6 μm $^6$LiF layer (thin converter) deposited onto a carbon fiber plate (a), coupled to a 300 μm thick silicon detector (b), and enclosed in a thin aluminum box (d). Right: SiLiF64 made of a double sandwich of four 16 μm $^6$LiF layers (thick converters) deposited onto carbon fiber plates (c), and enclosed in a thin aluminum box (d).

The silicon detectors we chose for our purpose are fully depleted, double sided, and have 3 cm x 3 cm x 300 μm size in order to have a reasonably large area and a comfortable detector capacitance of 300 pF, thus implying moderately low cost and no need of high performance electronics.

In Fig. 2 we show one of the calibrated detectors during the measurement in front of the Thermal Neutron Calibration Facility. The light beams from the laser alignment system are visible. The detectors were biased at 30 V, so that the silicon diodes were fully depleted. This was especially important for the SiLiF64, as the neutron converters were installed on both faces of each silicon diode and the full depletion regime is mandatory to get the same response from the front and the back sides. The operating principle is quite simple: we choose a neutron discrimination energy threshold and declare as neutrons all the counts above that threshold. Below the threshold there will be other neutron events mixed with gamma ray events and background noise. Of course the higher the threshold the cleaner the neutron signal will be, but at the same time the lower the detection efficiency will be. Therefore one has to find a trade-off between purity and efficiency, which may depend on the specific application and on the user needs.

## 3 Simulation procedure

The detectors were simulated by means of the well known GEANT4 toolkit ([18],[19]), including the aluminum box, the carbon fiber substrate, the $^6$LiF neutron converter, and the silicon diode. The supporting printed circuit board, the cable and the connector were not included in the simulation. $2 \times 10^6$ monoenergetic neutrons were generated for each case, with 25.3 meV kinetic energy, uniformly and perpendicularly irradiating the detector area. A preliminary set of simulations was run in order to decide the optimal thickness for the $^6$LiF converter. We simulated a configuration made of a single converter layer placed in front of the silicon detector, like in Fig. 1 left, but with the converter thickness varying from 8 to 16 μm in steps of 1 μm. The resulting distributions, reported in Fig. 3, show the typical evolution of the deposited energy spectrum shape as the converter thickness increases. For each simulated converter thickness the detection efficiency was evaluated as a function of the discrimination threshold, and the resulting plots are shown in Fig. 4. In Fig. 5 we show the detection efficiency as a function of the converter thickness when the discrimination threshold is set at 1.5 MeV, that is the safest threshold value, as will be shown in section 4. The plot shows that the detection efficiency starts to saturate with a converter thickness around 15 μm. Moreover, we have seen that around 20 μm the $^6$LiF deposited layer starts to lose mechanical stability. This is why we fixed our standard $^6$LiF layer thickness at 16 μm.

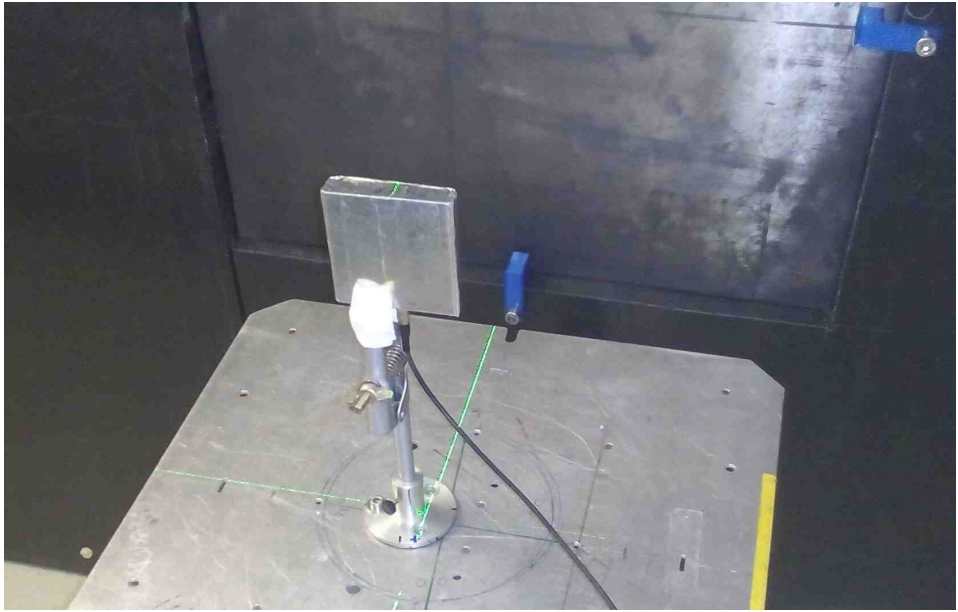

Fig. 2. One of the calibrated detectors during the measurement in front of the Thermal Neutron Calibration Facility. The light beams from the laser alignment system are visible.

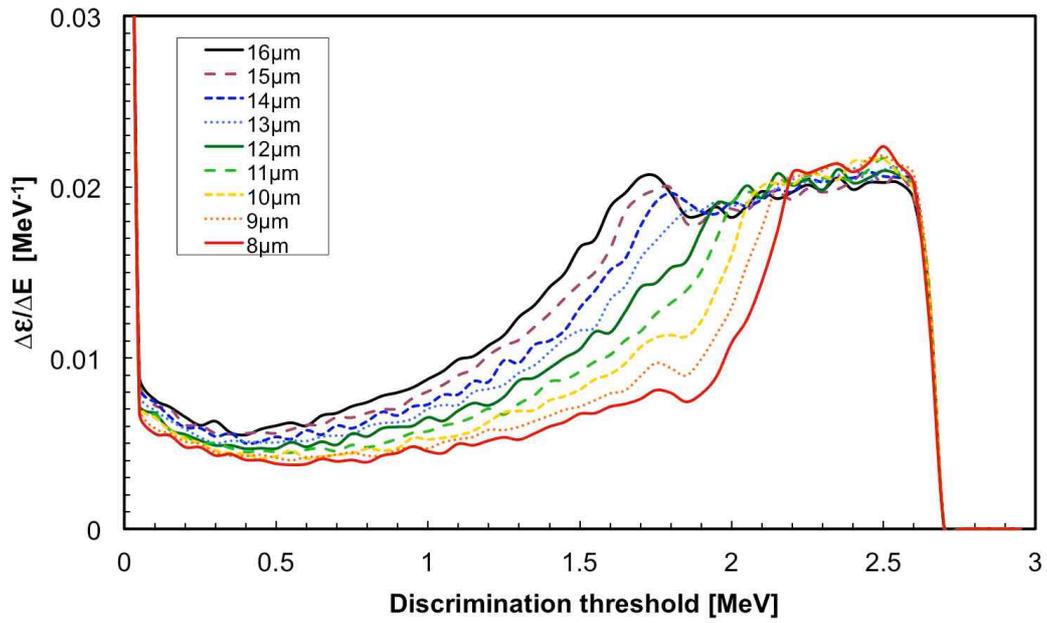

Fig. 3. Simulated energy spectrum (counting rate ε normalized to the unit flux) for several thicknesses of a single $^6$LiF converter placed in front of the silicon detector.

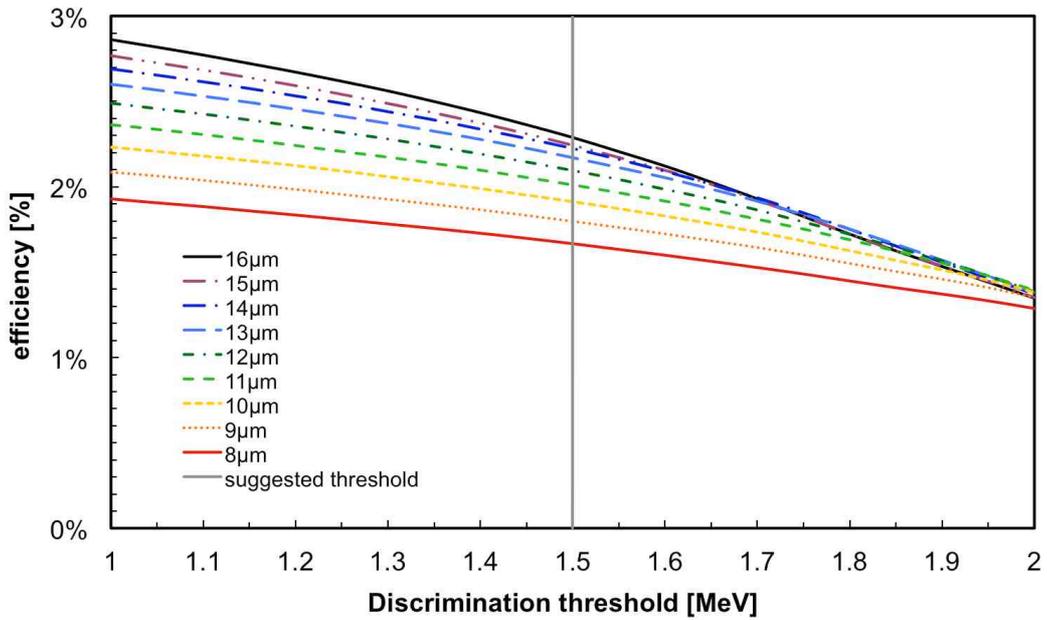

Fig. 4. Detection efficiency as a function of the discrimination threshold for each simulated converter thickness.

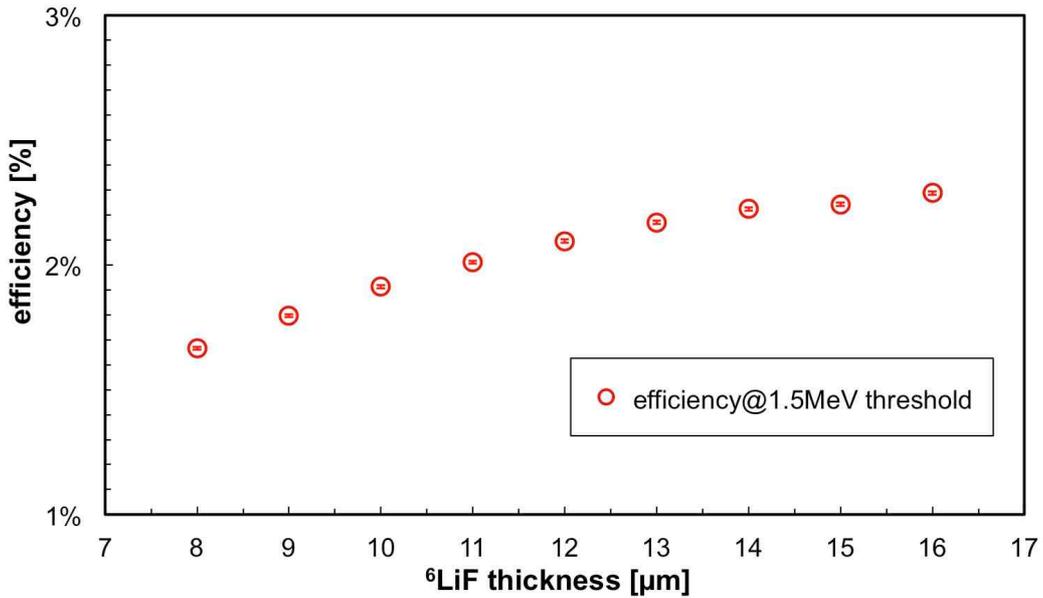

Fig. 5. Detection efficiency as a function of the converter thickness with the discrimination threshold set at 1.5 MeV.

## 4  Measurement results

The triton and the alpha particle emitted from the reaction (1) following a neutron capture have well defined energies, which are degraded after crossing the residual converter layer thickness and the thin air layer between the converter face and the silicon detector. Therefore, the thinner the converter the more accurate is the shape of the energy distribution measured by the silicon detector. Indeed, by looking at Fig. 6 where we reported the energy distribution (counting rate $\varepsilon$ normalized to the unit flux) as a function of the energy, one can see that using a 1.6 μm $^6$LiF converter the triton and alpha contributions to the spectrum are well identified. The alpha and triton endpoint energies are easily spotted on the spectrum, and this allows a perfect energy calibration of the detector, quite useful to set the neutron discrimination threshold with high precision. The agreement with the simulated spectrum is remarkable, considering that the experimental data were normalized to the nominal flux and therefore both spectra are in absolute units.

Unfortunately with such a thin converter the neutron detection efficiency is of the order of 0.5 %, quite low to allow a realistic employment of this detector. However, due to its spectral features and precision of its efficiency calibration, it can be quite useful as a reference for the efficiency calibration of other detectors. In Fig. 7 we show the energy spectrum as measured with the SiLiF64 detector, along with the simulation result, normalized to the unit flux. The agreement between simulation and measurement is remarkable also in this case, thus validating both the simulation tool and the detector behavior. In Fig. 8 we reported the energy spectra as measured by the SiLiF64 with and without the cadmium plate installed on the neutron source exit window. One can immediately see that on the one hand the neutron contribution decreases by about two orders of magnitude, whereas the gamma contribution is relevant at least up to 1÷1.5 MeV.

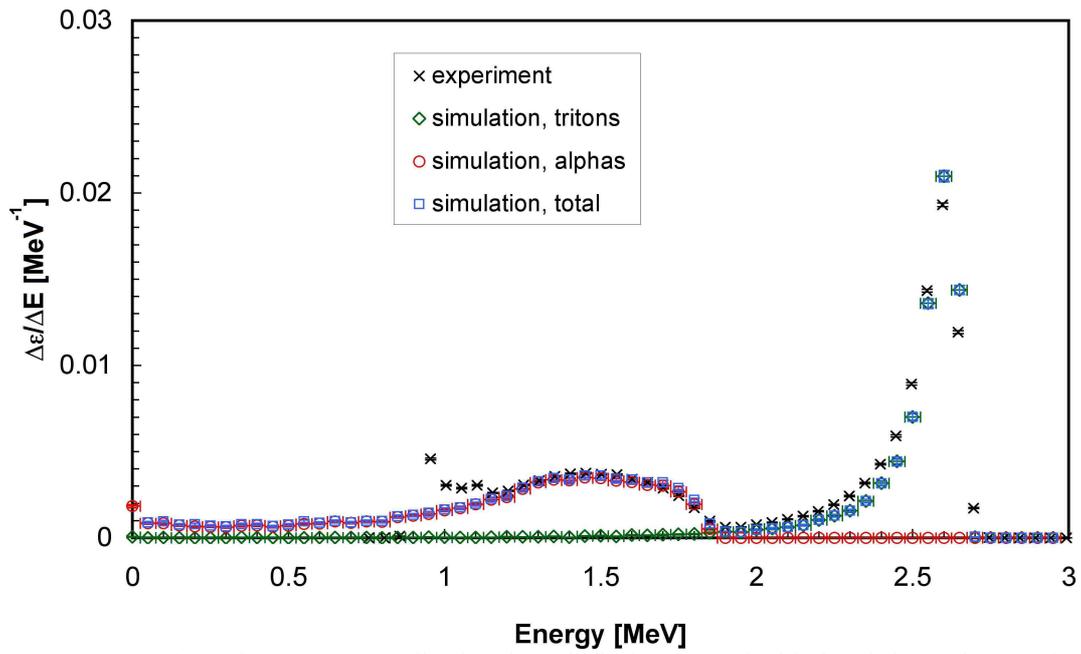

Fig. 6. Energy spectrum (counting rate ε normalized to the unit flux) measured with the SiLiF1.6 detector (crosses) and compared with the simulated one (squares). Also reported are the separate simulated contributions from alphas (circles) and tritons (diamonds).

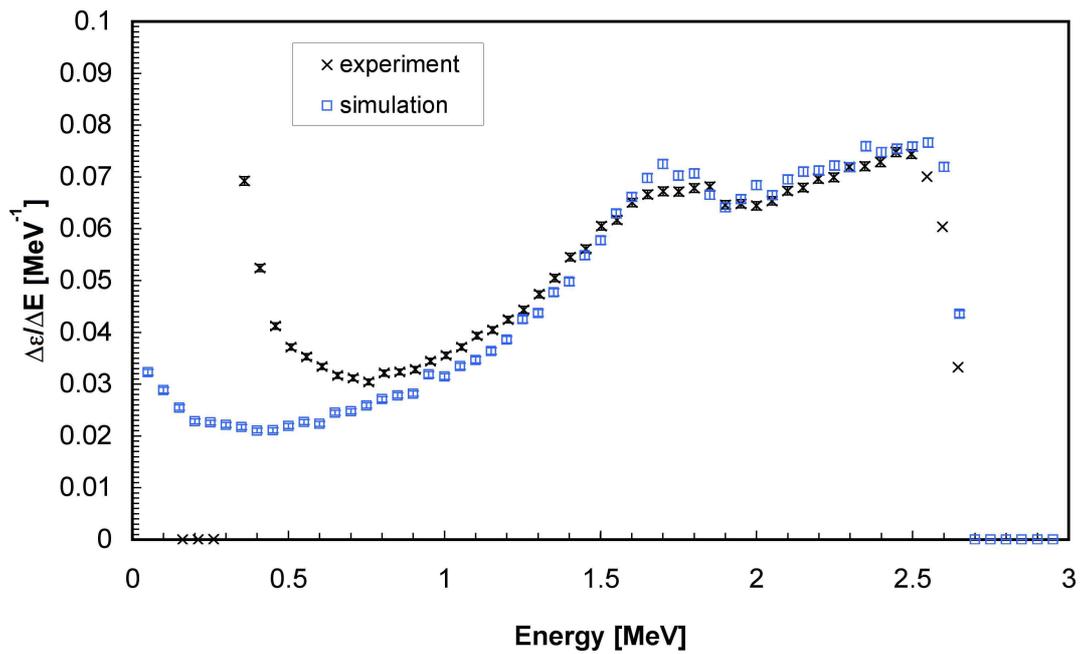

Fig. 7. Energy spectrum (counting rate ε normalized to the unit flux) measured with the SiLiF64 detector (crosses) and compared with the simulated one (squares).

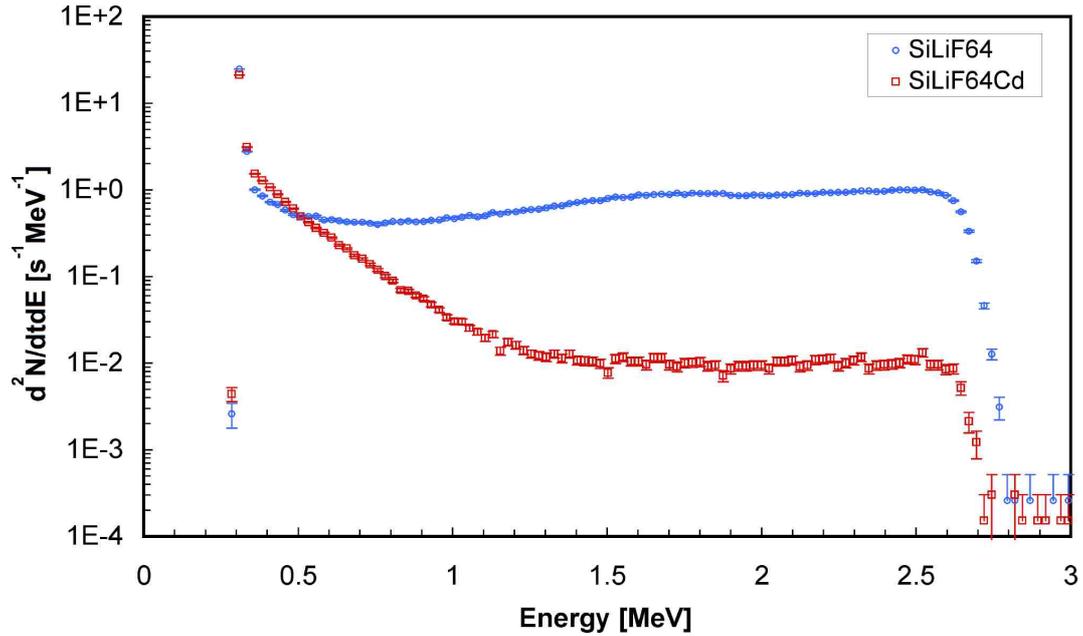

Fig. 8. Energy spectrum measured with the SiLiF64 detector, with and without the cadmium plate.

In Fig. 9 we reported the ratio between the two spectra of Fig. 8. In a pure neutron field such a ratio would be flat, whereas the presence of gamma rays, whose flux increases when the cadmium plate is inserted, modifies its shape. The plot shows rather clearly that in order to have a good γ/n discrimination the neutron discrimination threshold value should be chosen at 1.5 MeV: the gamma contribution decreases with increasing detection threshold, and above 1.5 MeV the ratio becomes flat as expected. However, lower threshold values can be safely employed in applications where there are no high energy gamma rays (as a reference, the γ/n contamination probability from $^{60}$Co gamma rays when setting a 1.5 MeV discrimination threshold is ≤ $10^{-12}$ [11]).

The corresponding plots for the SiLiF1.6 detector shown in Fig. 10 and Fig. 11 confirm, even though with a lower statistics, the behavior of Fig. 9 and the choice of 1.5 MeV as neutron discrimination threshold value.

In Table 1 we listed the simulated and measured efficiency for the two detectors. The measured data have a low statistical uncertainty, whereas the systematic one is more relevant, especially on the SiLiF64 detector, because of the energy calibration. Indeed, while the energy calibration of the SiLiF1.6 can be quite precise due to the presence of the two very well defined endpoints for alphas and tritons, this is not the case for a thicker converter layer, as can be easily seen by comparing Fig. 6 and Fig. 7. In both cases the triton endpoint energy can be identified much more precisely than the alpha one, which thus dominates the energy calibration uncertainty.

Therefore the systematic uncertainty was estimated by assuming a ±10 keV uncertainty in the alpha endpoint of the SiLiF1.6 and a ±50 keV in the alpha endpoint of SiLiF64. The results look satisfactory with a quite good agreement between data and simulations. This gives us confidence on the reliability of the simulation tool and of the detectors in view of new designs.

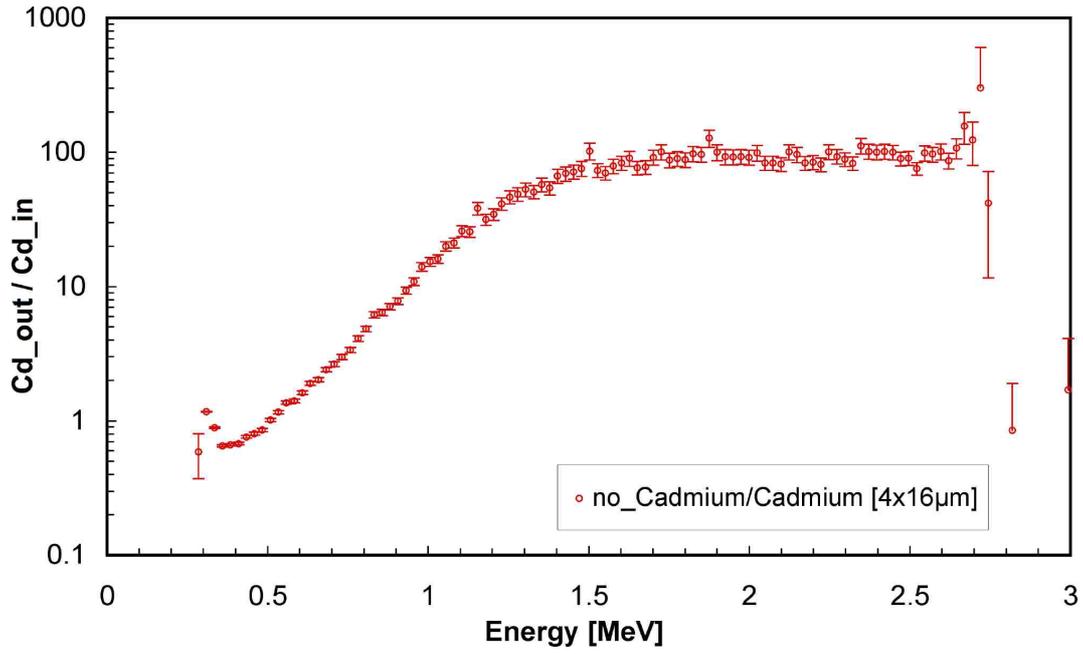

Fig. 9. Ratio between the two spectra of Fig. 8, that justifies the choice of 1.5 MeV as neutron discrimination threshold value.

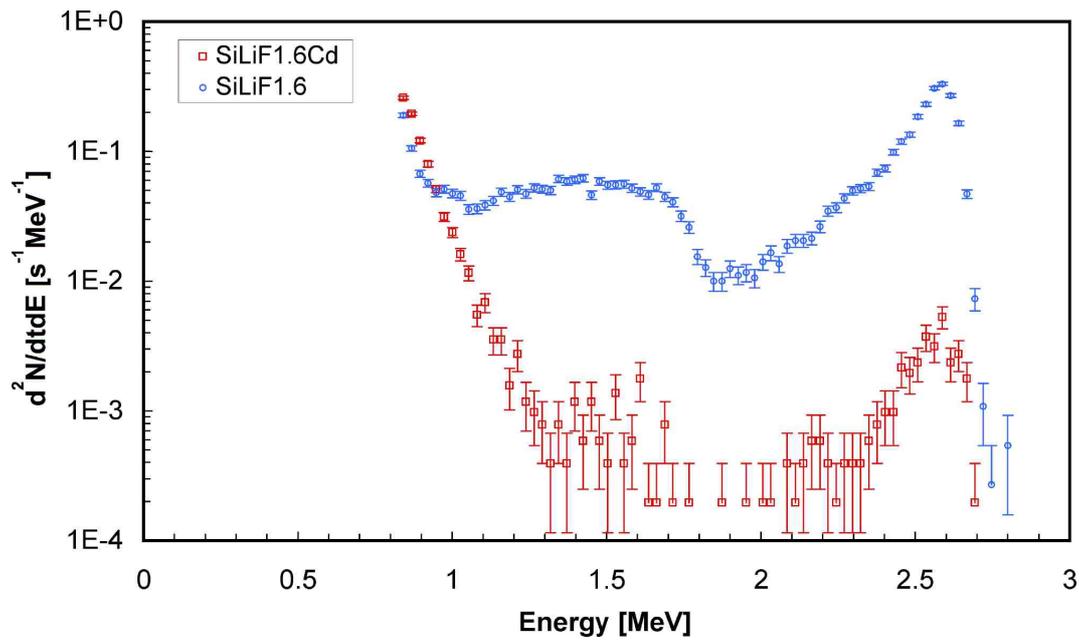

Fig. 10. Energy spectrum measured with the SiLiF1.6 detector, with and without the cadmium plate.

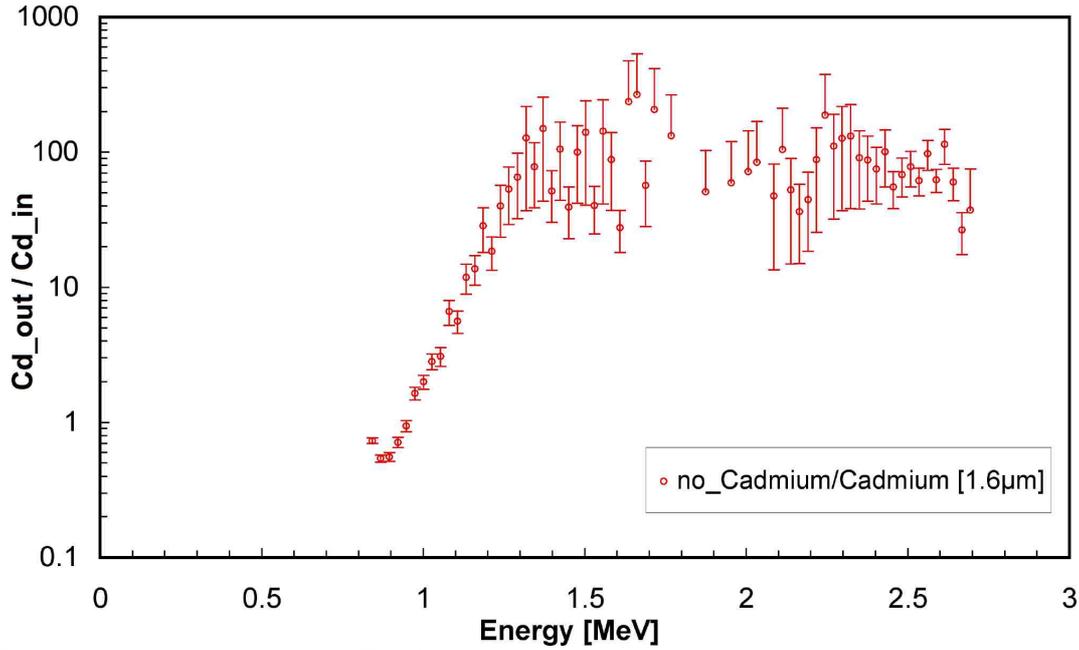

Fig. 11. Ratio between the two spectra of Fig. 10, that even with a lower statistics confirms the behavior of Fig. 9 and the choice of 1.5 MeV as neutron discrimination threshold value.

Table 1 - Simulated and measured detection efficiency

| detector | simulated efficiency | measured efficiency |
|---|---|---|
| SiLiF1.6 | 0.48 % | 0.50 ± 0.02 (syst) % |
| SiLiF64 | 8.25 % | 7.95 ± 0.35 (syst) % |

## 5 Perspectives

These neutron detectors have also been tested at neutron beam facilities like ISIS (Rutherford Appleton Laboratory, UK), and n_TOF (CERN) where a few samples are already in operation for beam flux and 2D profile measurements. Forthcoming applications are foreseen for the online monitoring of spent nuclear fuel casks in interim storage sites, and to this aim a new set of simulations for several different configurations has already been started, along with the planning of additional efficiency calibrations using the same facility at PTB.

## 6 Conclusion

In order to cope with the need of new thermal neutron detectors as an alternative to $^3$He tubes in research, industrial, safety and homeland security applications, we have developed, tested and characterized solid state silicon detectors coupled with $^6$LiF neutron converters. Even though their detection efficiency is only around 8 %, the gamma rejection performance and the rather low cost as compared to $^3$He tubes, make these detectors quite interesting for several applications, especially those with continuous monitoring purposes.

## 7 Acknowledgments

We thank Sebastian Fässer for his invaluable support during the measurements at PTB. We are strongly indebted with Carmelo Marchetta and Antonio Massara for the preparation of the $^6$LiF converters. We are grateful to Andreas Zimbal for his suggestions and indications. A heartly felt acknowledgment is due to Marco Ripani for his constant encouragement and support within the framework of the INFN-E Strategic Project.

# 8 References

[1] D.Henzlova et al., "Current Status of 3He Alternative Technologies for Nuclear Safeguards", NNSA USDOE and EURATOM, LA-UR-15-21201.

[2] R. T. Kouzes, "The 3He supply problem", Tech. Rep. PNNL-18388, Pacific Northwest National Laboratory (Richland, WA, 2009).

[3] R.T. Kouzes et al., "Neutron detection alternatives to 3He for national security applications", Nucl. Instrum. Methods Phys. Res. A623 (2010) 1035.

[4] N. Ensslin, in "Passive Nondestructive Assay of Nuclear Materials", NUREG/CR-5550 LA-UR-90-732, 457 (1991)

[5] P. Finocchiaro, "Radioactive Waste: A System for Online Monitoring and Data Availability", Nucl. Phys. News v.24, n.3, (2014) 34

[6] P.Finocchiaro, "DMNR: a new concept for real-time online monitoring of short and medium term radioactive waste", in Radioactive Waste: Sources, Types and Management, Nova Science Publishers, (2011)

[7] A. Pappalardo et al., "Characterization of the silicon+6LiF thermal neutron detection technique", Nucl. Instrum. Methods Phys. Res. A810 (2016) 6.

[8] D. S. McGregor et al., "Design considerations for thin film coated semiconductor thermal neutron detectors—I: basics regarding alpha particle emitting neutron reactive films", Nucl. Instrum. Methods Phys. Res. A500 (2003) 272.

[9] M. Barbagallo et al., "Thermal neutron detection using a silicon pad detector and 6LiF removable converters", Rev. Sci. Instrum. 84, (2013) 033503.

[10] http://www.radectech.com/

[11] S. Lo Meo, L. Cosentino, A. Mazzone, P. Bartolomei, P. Finocchiaro, "Study of silicon+$^6$LiF thermal neutron detectors: GEANT4 simulations versus real data", Nucl. Instr. Meth. A866 (2017) 48.

[12] L.Cosentino et al., "Silicon detectors for monitoring neutron beams in n-TOF beamlines", Rev. Sci. Instrum. 86 (2015) 073509.

[13] A.Pappalardo, C. Vasi, P. Finocchiaro, "Direct comparison between solid state Silicon+6LiF and 3He gas tube neutron detectors", Results in Physics 6 (2016) 12.

[14] F.Gunsing et al., "Status and outlook of the neutron time-of-flight facility n_TOF at CERN", Nucl. Instr. Meth. B261 (2007) 925.

[15] E. Chiaveri et al., "Proposal for n_TOF Experimental Area 2", CERN-INTC-2012-029 / INTC-O-015 09/

[16] M. Luszik-Bhadra, M. Reginatto, H.Wershofen, B.Wiegel, and A. Zimbal, "New PTB Thermal Neutron Calibration Facility: first results", Radiation Protection Dosimetry (2014), Vol. 161, No. 1–4, pp. 352–356.

[17] M. Luszik-Bhadra, D. Radeck, M. Reginatto, H. Wershofen, M. Zboril, and A. Zimbal, "The PTB Thermal Neutron Calibration Facility", PTB Report, Physikalisch-Technische Bundesanstalt, Braunschweig, to be published.

[18] S.Agostinelli et al., "Geant4 - a simulation toolkit", Nucl. Instrum. Methods Phys. Res. A506 (2003) 250.

[19] S. Lo Meo et al., "GEANT4 simulations of the n TOF spallation source and their benchmarking", Eur. Phys. Jou. A51 (2015) 160.